\documentclass[twocolumn,preprintnumbers,amsmath,amssymb]{revtex4-1}

\usepackage[dvips]{graphicx}
\usepackage{epsfig}
\usepackage{dcolumn}
\usepackage{bm}
\usepackage[T1]{fontenc}       

\usepackage{color}
\usepackage{float}
\usepackage[colorlinks]{hyperref}

\begin{document}

\title{Pattern formation mechanisms in sphere-forming diblock copolymer thin films}

\author{Leopoldo R. G\'omez}
\affiliation{Instituto de F\'isica del Sur (IFISUR), Consejo Nacional de Investigaciones Cient\'ificas y
T\'ecnicas (CONICET),  Universidad Nacional del Sur, 8000 Bah\'ia Blanca, Argentina}

\author{Nicol\'as A. Garc\'ia}
\affiliation{Instituto de F\'isica del Sur (IFISUR), Consejo Nacional de Investigaciones Cient\'ificas y
T\'ecnicas (CONICET),  Universidad Nacional del Sur, 8000 Bah\'ia Blanca, Argentina}

\author{Richard A. Register}
\affiliation{Department of Chemical and Biological Engineering, Princeton University, Princeton, New Jersey, 08544, USA}

\author{Daniel A. Vega}
\affiliation{Instituto de F\'isica del Sur (IFISUR), Consejo Nacional de Investigaciones Cient\'ificas y
T\'ecnicas (CONICET),  Universidad Nacional del Sur, 8000 Bah\'ia Blanca, Argentina}

\date{\today}

\begin{abstract}
The order-disorder transition of a sphere-forming
block copolymer thin film was numerically studied through a Cahn-Hilliard model.
Simulations show that the fundamental mechanisms of pattern formation are spinodal decomposition and nucleation and growth. The range of validity of each relaxation process is controlled by the spinodal and order-disorder temperatures. The initial stages of spinodal decomposition are well-approximated by a linear analysis of the evolution equation of the system. In the metastable region, the critical size for nucleation diverges upon approaching the order-disorder transition, and reduces to the size of a single domain as the spinodal is approached. Grain boundaries and topological defects inhibit the formation of superheated phases above the order-disorder temperature.
The numerical results are in good qualitative agreement with experimental data on sphere-forming diblock copolymer thin films.
\end{abstract}

\maketitle

\section{\label{sec:intro}Introduction}
Many practical applications of polymers and other soft
matter involve the self-assembly of the system into complex multidomain morphologies \cite{Hamley}.
The final properties and applications of such materials depend on the ability to control the
morphology by adjusting molecular features and macroscopic
variables. For example, by appropriate control over their molecular architecture, block copolymers can be designed as
rigid and transparent thermoplastics, or as soft and flexible
elastomers, depending on the features of their building blocks \cite{Hamley}.

The most important features of the block copolymer phase behavior are
already captured by the simplest A-B diblock architecture \cite{Hamley,fredicksonPT}. Here the
unfavorable interactions between blocks, and the
constraints imposed by the connectivity between their constituents, results in
a nanophase separation leading to the formation of periodic morphologies.
For example, depending on molecular size, temperature, and relative volume fraction of the two blocks, diblock copolymer melts can develop body centered-cubic arrays of spheres, hexagonal patterns of cylinders, gyroids or lamellar structures. For these systems, the periodicity of the self-assembled pattern is mainly controlled by the average molecular weight, and typically is in the range of 10-100 nm. Also, since the magnitude of the interblock repulsive interaction generally
diminishes with temperature, an order-disorder transition can
be induced thermally at the order-disorder transition
temperature $T_{ODT}$ \cite{Hamley,fredicksonPT}.

\begin{figure*}
\begin{center}
\includegraphics[width=14cm]{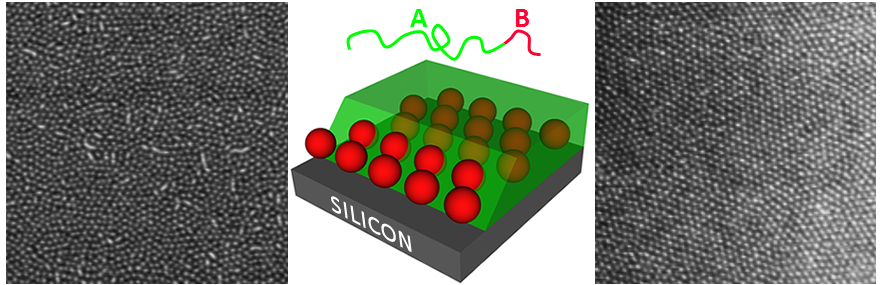}
  \caption{AFM phase images of the block copolymer thin film supported on a silicon substrate  after annealing at T = 333 K. Left and right panels show the pattern configuration after 5 minutes and 255 minutes of thermal annealing, respectively. Image size: 1.0 $\mu$m x 1.0 $\mu$m. The scheme of the central panel shows the block copolymer configuration in the thin film. Here the minority phase (polystyrene) forms hexagonally-packed arrays of nearly spherical domains.}
 \label{fig:fig1}
\end{center}
\end{figure*}

During recent decades, the properties of self-assembling copolymers have received
great attention because of their potential use in nanotechnology
\cite{HamleyNanotech,Lodge,LeiberNatMat,Segalman,Darling,Kim,CBates}. Applications of block copolymer systems include, among others, templates for nanoporous materials, solar cells, and  photonic crystals.
Perhaps the most pressing application for understanding pattern formation in two-dimensional thin film
systems is block copolymer lithography \cite{Segalman,Darling,Kim,CBates}. This process uses self-assembled
patterns, such as single layers of cylinders or spheres in copolymer thin films, as
templates to fabricate devices at the nanometer length scale.
However, the use of these materials to obtain lithographic masks
requires the production of structures with long-range order. Since
the degree of ordering is controlled mainly by the density of
topological defects, it is important to determine the physical
mechanisms involved in the nanophase separation process and the
effects of the thermal history on the pattern morphology.

Experimentally, the time evolution of the density of topological defects and correlation lengths at $T<T_{ODT}$ has been studied through Atomic Force Microscopy (AFM) in block copolymer thin films with different morphologies \cite{Harrison1,Harrison2002,Trawick,Marencic1,Marencic2,Harrison2}. By comparing the density of disclinations
(orientational defects) and the correlation length, it was shown
that in cylinders (smectic phase) the dominant mechanism of
coarsening involves mostly the annihilation of complex arrays of
disclinations \cite{Harrison1}. On the other hand, in monolayers of sphere-forming thin films
with hexagonal order it was found that the majority of the defects
are condensed in grain boundaries \cite{Harrison2}. The orientational correlation
length was found to grow following a power law, but with a higher
exponent than the translational correlation length \cite{Harrison2,VegaPRE}.
However, as the typical exponents observed in the scaling laws of these experimental systems are relatively small ($\sim \frac{1}{5}$  for the translational correlation length and $\frac{1}{4}$ for the orientational correlation length), they are hard to distinguish from logarithmic, glass-like, dynamics \cite{GomezPRL}.

The process of phase separation and the kinetics of ordering in different two-dimensional systems have been studied through different phase field models \cite{VegaPRE,GomezPRL,GarciaPRE,Shimacromol}. In particular, simulations of hexagonal systems  with a Cahn-Hilliard model were found to be in good agreement with
experimental data for block copolymer thin films. Through simulations it was also shown that the
orientational correlation length grows via annihilation of
dislocations \cite{VegaPRE}. In addition, simulations have also shown that triple points, regions
where three grains meet, control the dynamics of defect annihilation  and
can lead to the formation of metastable configurations of domains that slow down the dynamics, with correlation lengths growing logarithmically
in time \cite{GomezPRL}.

In block copolymer systems the degree of order and content of defects depend not only on
$T_{ODT}$, but also on thermal history, depth of quench, and
other characteristic temperatures, like the glass transition
temperature for each block, the spinodal temperature, and temperature of crystallization (if any) \cite{Hamley,fredicksonPT}.

In general, in first-order phase transitions the depth of
quench determines whether the system relaxes to equilibrium by means of
spinodal decomposition or nucleation and growth \cite{ChaikinBook,DebenedettiBook}. Spinodal
decomposition is the relaxation process of a system quenched
into an unstable state. In this case the phase transition is a
spontaneous process, beginning with the amplification of fluctuations which are small in amplitude but large in extent. Nucleation and growth
is the physical mechanism of relaxation emerging when the system is quenched into a metastable state. This
is an activated process, and a free energy barrier must be overcome in
order to relax to the stable phase. In polymeric systems both mechanisms can be inhibited due to the crystallization or vitrification of any of the copolymer building blocks \cite{Hamley}.

In this paper we  study the disorder-order transition and the
ordering kinetics in  sphere-forming block
copolymer thin films as a function of the depth of quench. The dynamics are
studied through a Cahn-Hilliard equation and compared with experiments on diblock copolymer thin films.
We have organized the paper as follows: Section II presents the experimental system.  In section III we
present the equations of evolution of the order parameter for diblock copolymer systems and the classical
linear instability analysis  of the Cahn-Hilliard equation. Section
III.i describes the numerical scheme employed to solve the model equations.
Results and concluding remarks are presented in sections IV and V, respectively.

\section{Experimental System}
\label{sec:Exps}
The sphere-forming diblock copolymer used in this work consists
of a chain of polystyrene (PS, 3300 g/mol) covalently bonded to a chain of poly(ethylene-alt-propylene)
(PEP, 23100 g/mol). The copolymer was synthesized through sequential living anionic polymerization
of styrene and isoprene followed by selective saturation of the isoprene block \cite{Sebastian}.  Because the two polymer species are
immiscible, the minority polystyrene forms spherical microdomains within the
majority poly(ethylene-alt-propylene). The bulk morphology of the PS-PEP diblock copolymer consists of arrays of spherical domains of PS packed with a BCC order, with an order-disorder temperature
(ODT) of  $394 \pm 2$ K, according to small-angle X-ray scattering experiments.
The glass transition temperature for the PS block was estimated to be $T_g^{PS}\sim 320$ K while the glass transition temperature of the PEP block is well below room temperature \cite{Angelescu}.

The block copolymer was deposited (film thickness ca. 25 nm) on silicon substrates via
spin coating from a disordered state in toluene, a good solvent for both blocks, to produce
a quasi-2-dimensional periodic hexagonal lattice of polystyrene spheres within a matrix of poly(ethylene-alt-propylene). Order was induced through vacuum annealing above the glass transition temperature of both blocks and below
the $T_{ODT}$ of the block copolymer \cite{Fasolka}.

Samples were imaged with a Veeco Dimension 3000 AFM in tapping mode, using phase contrast imaging. The contrast is provided by the
difference in elastic modulus between the hard polystyrene spheres and the softer poly(ethylene-alt-propylene) blocks. The repeat spacing for the block copolymer is $25$ nm (as measured on thin films by AFM). The spring constant of the tip (uncoated Si) was $\sim 40$ N/m and its resonant frequency $300$ kHz.

During spin coating, most of the toluene evaporates rapidly (within a few seconds), so the block copolymer thin film suffers a relatively quick quench well below the glass transition temperature of the PS domains, inhibiting the relaxation of the early nanophase separated structure towards equilibrium.  The  deep and quick quench below the  spinodal  and order-disorder temperatures forms a nanodomain structure that contains a high density of defects.  Figure 1 shows tapping-mode AFM phase images of the PS-PEP system at a very early stage of annealing. Note in this figure the large content of defects and that the spherical domains are not well defined.   On the other hand, after $\sim 4$ hours of annealing above the glass transition temperature of the PS block, the pattern show a higher degree of order and a well-defined structure of spherical domains.
\begin{figure}[b]
\begin{center}
        \includegraphics[width=7.8cm]{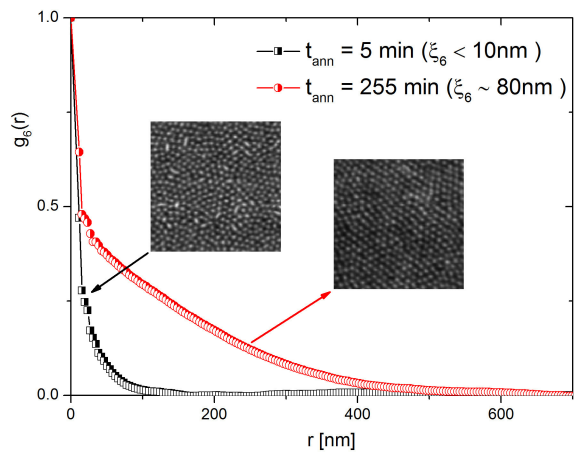}
    \caption{ Orientational correlation function $g_6(r)$ for the two patterns shown in Figure 1. As annealing time increases, there is an increasing order in the system, as shown by the slower decay in $g_6(r)$ with $r$ and the increasing in the correlation length $\xi_6$.}
    \label{fig:correxp}
    \end{center}
\end{figure}
In order to better quantify the degree of order in this system, here we calculate the orientational order parameter $\xi_6$.
Using standard image tools to identify the position of the individual spheres and applying a Delaunay triangulation \cite{Harrison2,Trawick} it is possible to determine the inter-sphere bond orientation $\theta(r)$, with regard to a reference axis and then to evaluate the local orientational order parameter at a
position $\textbf{r}$: $\Phi(\textbf{r})=\exp[6i(\theta(\textbf{r}_1)-\theta(\textbf{r}_2))]$, where $\textbf{r}=\textbf{r}_1-\textbf{r}_2$ \cite{VegaPRE,GomezPRL}. The azimuthally-averaged correlation function  $g_6(r)=\langle\Phi(\textbf{r})\rangle$ was then calculated
and the correlation length $\xi_6$ was measured by fitting $g_6(r)$
with an exponential $\exp(-r/\xi_6)$.
Figure 2 shows the azimuthally-averaged correlation function  $g_6(r)$ for the two pattern configurations shown in Fig. 1.  During annealing, the correlation length of the film increases from $\lesssim 10$ nm
to $80$ nm upon increasing the annealing time from 5 to 255 minutes.

\section{Model}
\label{sec:Model}
The phase transition and the dynamics of nanophase
separation for a diblock copolymer can be described through the Cahn-Hillard model
\cite{Hamley}. A convenient order parameter for the diblock
copolymer can be defined in terms of the local volume fractions
for each block as $\psi = \phi_A - \phi_B - (1-2f)$. Here $\phi_A$
and $\phi_B$ are the local densities for the \textit{A} and \textit{B}
blocks, respectively, and $f$ is the volume fraction of one block in the copolymer. The dynamics can be described by the following time-dependent equation for a conserved order parameter \cite{Karma}:
\begin{equation}
\label{ecu:CHC} \frac{\partial \psi}{\partial t}=M \, \nabla^2
\frac{\delta F\{\psi\} }{\delta \psi}
\end{equation}
In this equation $M$ is a phenomenological mobility coefficient,
and $F\{\psi\}$ is the free energy functional.

In the mean field approximation the free energy functional for a diblock copolymer
can be decomposed into a sum of short-range and long range terms \cite{Leibler,OthaKawasaki}:
\begin{equation}
\label{ecu:CHC1}
 F\{\psi\}=F_S\{\psi\}+F_L\{\psi\}
\end{equation}
The short range contribution $F_S$ has the form of a Landau free
energy and can be expressed as:
\begin{equation}
\label{ecu:CHC2}
 F_S\{\psi\}=\int d\textbf{r}
\{W(\psi)+\frac{1}{2}\,D\,(\nabla \psi)^2 \}
\end{equation}
where $W(\psi)$ represents the mixing free energy of the
homogeneous blend of disconnected \textit{A} and \textit{B} homopolymers, the term
containing the gradient represents the free energy penalty
generated by the spatial variations of $\psi$  (interfacial energy) and
$D$ is a parameter related to the segment length. The free energy $W(\psi)$) has the
form of a non-symmetrical double well:
\begin{equation}
W(\psi )= -\frac{1}{2} [\tau - a (1- 2 f)^2] \, \psi^2+
\frac{v}{3} \, \psi ^3 + \frac{u}{4} \, \psi^4
\end{equation}
Here the parameter $\tau$ is related to temperature by means of
the Flory-Huggins parameter $\chi$ through \cite{RenHamley}:
\begin{equation}
\tau=8\,f\,(1-f)\,\rho_0\,\chi-\frac{2\,s(f)}{f\,(1-f)\,N}
\end{equation}
where $\rho_0$ is the monomer density, $N=N_A+N_B$ is the total
number of monomers in the diblock copolymer chain, and $s(f)$ is a constant of
order unity. In this work we consider the parameters $a$, $v$, and
$u$ as phenomenological constants.

The term $F_L\{\psi\}$, representing the long-range contribution that accounts for the connectivity between blocks, can be expressed as \cite{OthaKawasaki}:
\begin{equation}
\label{ecu:CHC3} F_L\{\psi \}=\frac{b}{2} \int d \textbf{r}_1 \int
d \textbf{r}_2 \,G(\textbf{r}_1 -
\textbf{r}_2)\,\psi(\textbf{r}_1) \,\psi(\textbf{r}_2)
\end{equation}
where $G$ is the solution of $\nabla^2\,G(\textbf{r})=-\delta
(\textbf{r})$.
Then, using equations \ref{ecu:CHC1}-\ref{ecu:CHC3},  eqn.
\ref{ecu:CHC} takes the form:
\begin{equation}
\label{ecu:CHCfinal}
\frac{1}{M} \frac{\partial \psi}{\partial t}=\nabla^2(\tilde{f}(\psi) -D\nabla^2 \psi)-b \psi
\end{equation}
where $\tilde{f}(\psi)$ is given by:
\begin{equation}
\tilde{f}(\psi)=-[\tau-a\, (1-2\,f)^2]\,\psi + v\,\psi^2+u\,\psi^3
\end{equation}
\begin{figure}[t]
\begin{center}
    \includegraphics[width=7.6cm]{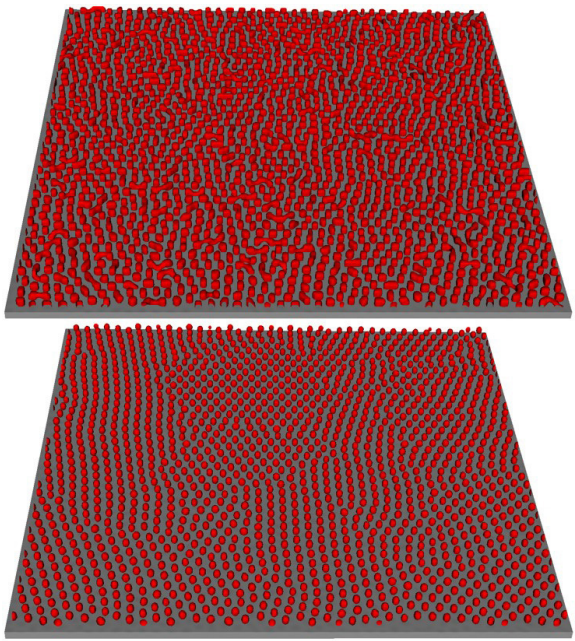}
    \caption{ Phase separation process after $4.5\times10^4$ time steps for a sphere-forming diblock copolymer monolayer quenched within the spinodal region. Top and bottom panels correspond to $\tau_r=\tau / \tau_{s}-1=0.34$ and $\tau_r=0.025$, respectively.  In order to describe the structure of domains developed by the minority phase, here we employ a contour plot at a fixed value of the order parameter ($\psi=-0.22$).}
     \label{fig:cds3Da}
\end{center}
\end{figure}
Equation \ref{ecu:CHCfinal} is actually a coarse grain phase field model. In the last years these kind of models have been used to study a variety of system under different conditions \cite{ProvatasElder}, between others, shear and other external fields, curvature and patterned substrates, and pattern formation in 3D \cite{An-Chang-Shi, GomezVega3D,Pan,Huang,Zvelindovsky}.
\subsection{Numerical methods}
\label{sec:CDS}
In this work eqn. \ref{ecu:CHCfinal} was numerically solved under confinement between impenetrable walls that represent the interfaces with the air and the silicon wafer. To describe the confinement interactions we consider an external field that couples with the two components of the block copolymer system (see details in Ref. \cite{Abate2016}). The film thickness was fixed at the monolayer value.  We employed periodic boundary conditions along the (x,y)-axis.
 The evolution equation, eqn. \ref{ecu:CHCfinal}, was solved through the cell
dynamics method, which has been extensively used in the
non-equilibrium studies of this kind of system \cite{RenHamley}.
One of the main advantages of this model is that it is efficient over the time scales involved in the dynamics of defect annihilation, and is thus appropriate to describe the dynamics of coarsening (see Ref. \cite{Sevink} for an extensive and detailed analysis of the CDS method and a comparison with the Cahn-Hilliard-Cook model).
Another remarkable advantage of this model is its matricial nature (continuum discretization), which allows the implementation of a transparent parallelization into a GPU code. Here, we solved this model using a dual buffering scheme (Ping-Pong technique) and an optimized use of the different GPU memories \cite{Skadron,CUDA}.

Here we study how the process of pattern formation varies with the parameter related to the temperature $\tau$, while keeping the rest of the parameters fixed at the following values: $M=1.0$, $D=0.1$, $a=1.5$, $v=2.3$, $u=0.38$ and $b=0.01$. The initial homogeneous (disordered) state is simulated by a random noise
distribution of the order parameter.

Figure \ref{fig:cds3Da} shows two snapshots of the typical pattern configurations, obtained through simulations of monolayers,  observed at different conditions of annealing when an initially disordered system is quenched into the spinodal region ($\tau_s <\tau<\tau_{ODT}$).

\begin{figure}[t]
\begin{center}
    \includegraphics[width=7.5cm]{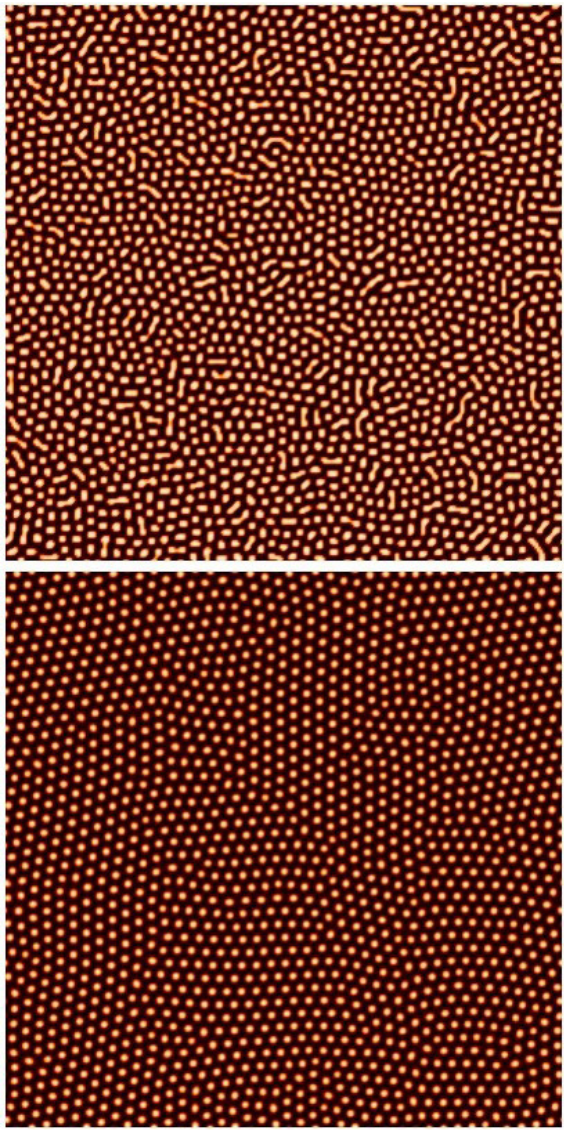}
    \caption{Two-dimensional pattern configurations for the 3D simulation data shown in Fig. \ref{fig:cds3Da}. Observe the similarities between these patterns and
those in the experimental system, shown in Fig. \ref{fig:fig1}.}
\label{fig:cds2D}
\end{center}
\end{figure}

To obtain a better comparison with the experimental
AFM phase images shown in Fig. 1, rather than considering
a contour plot to describe the individual PS spheres,
in Fig. 4 we integrate the order parameter describing the
density fluctuations along the direction perpendicular to
the substrate. In this way, the interfaces are smoothed
out, facilitating the comparison with AFM. Note the
qualitative agreement between experimental data (Fig. \ref{fig:fig1}) and numerical
results shown in Fig. \ref{fig:cds2D} for systems at different annealing conditions.

\subsection{Linear Instability Analysis}
Due to the strong degree of confinement in the monolayer, in general the formation of ordered structures and patterns can be well approximated as a 2D process.  In this section we consider temperatures in the vicinity of the order-disorder
transition ($T \lesssim T_{ODT}$) and study the pattern evolution of a
modulated structure in the plane $z=0$, whose order parameter profile can be
described by the following sum of Fourier modes:
\begin{equation}
\label{ecu:linstab}
\psi(\textbf{r},t)=\sum_\textbf{k} \psi_\textbf{k}\,exp \,{(i \,
\textbf{k}\cdot \textbf{r}+\lambda \, t)}
\end{equation}
where $\psi_\textbf{k}$ is the Fourier coefficient at $t = 0$ (we consider that $\textbf{k}$ is restricted to the thin film's middle plane and neglect the spatial variation of $\psi$ along the direction perpendicular to the thin film surfaces).

The stability of the solution $\psi = 0$ (high temperature phase) for
a system quenched into an unstable state can be studied by
linearizing equation \ref{ecu:CHCfinal} around $\psi = 0$ \cite{Vega2,Pezzutti}.
Substituting eqn. \ref{ecu:linstab} into the linearized equation
\ref{ecu:CHCfinal} it can be easily shown that the
amplification factor $\lambda$ satisfies:
\begin{equation}
\lambda(k)=- D\,k^4 + [\tau-a\,(1-2\,f)^2]\,k^2-b
\label{ecu:lambda}
\end{equation}
where $k=\|\textbf{k}\|$. The spinodal temperature $\tau_{s}$
can be calculated as the lowest value of $\tau$ for which an
extended mode can grow, i.e.,  $\lambda > 0$ for some $k$:
\begin{equation}
\label{ecu:ODT}
\tau_{s}=2\,\sqrt{D\,b}+a\,(1-2\,f)^2
\end{equation}
Thus, for  $\tau < \tau_{s}$, no extended mode can
be amplified and the state $\psi = 0$ is a stable or
metastable configuration. If $\tau > \tau_{s}$, some modes modes grow
exponentially with time. These growing modes are constrained according to:
\begin{equation}
\label{ecu:modospropagan}
k_1^2 \leq k^2 \leq k_2^2
\end{equation}
where:
\begin{equation}
\label{ecu:modospropagan2}
\begin{split}
{k_1^2=\frac{\Gamma - \sqrt{\Gamma^2-4\,D\,b} }{2\,D}} \\
k_2^2={\frac{\Gamma + \sqrt{\Gamma^2-4\,D\,b} }{2\,D}}
\end{split}
\end{equation}
with $\Gamma = \tau - a \,(1-2\,f)^2$. Moreover, the most
unstable mode (where $\lambda$ is maximal) is:
\begin{equation}
k_s=\sqrt{\frac{\Gamma}{2\,D}}
\end{equation}
if $\tau\to\tau_{s}$, this has the form $k_s\to(b/D)^{1/4}$,
and is the only mode which is unstable.

Note that, in general, the
spinodal temperature does not coincide with the disorder-order
temperature. If a homogeneous system is quenched between these
temperatures, it can remain in a metastable state, which can
relax only by nucleation and growth. For block copolymers the spinodal
and disorder-order temperatures coincide only for symmetric
lamellar morphologies\cite{Hamley}, $f=\frac{1}{2}$.
\begin{figure}[b]
\begin{center}
    \includegraphics[width=7.8cm]{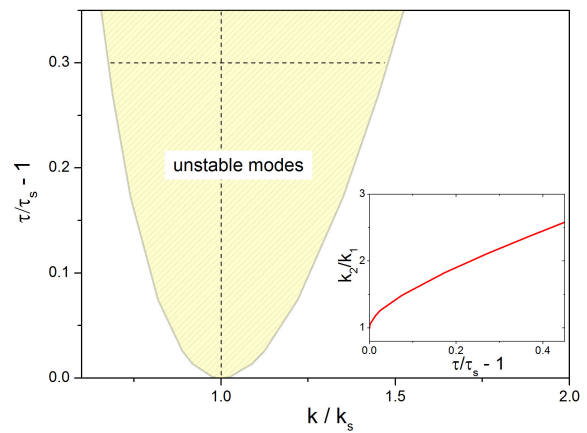}
    \caption{Range of unstable modes as a function of the reduced temperature. Inset: $k_2/k_1$ as a function of the reduced temperature. Note that the unstable modes are not evenly distributed around $k_s$.}
     \label{fig:unstablemodes}
\end{center}
\end{figure}

Figure 5 shows the range of unstable modes as given by eqn. \ref{ecu:modospropagan}. Note that the distribution of unstable modes is not symmetrically distributed around $k_s$, the wave vector at the spinodal. This skewness becomes progressively larger as the depth of quench increases. Consequently, for systems quenched in the neighborhood of the spinodal, one can expect a very strong length scale selectivity and a strong free energy penalty for elastic distortions which shift the system from the optimum $k_s$.  On the other hand, deep quenches stabilize the presence of different elastic distortions and thus, more disordered patterns can be expected. In addition, due to the skewness of the distribution, phase-separated systems generated  at deep quenches are characterized by an average wave vector $k>k_s$.

It is known that the linear instability analysis describes
only the initial stage (short times) of the spinodal decomposition
mechanism. As time proceeds, the dynamics become highly nonlinear
and higher-order wave numbers emerge \cite{Cahn,Langer}. We will discuss these other stages
of spinodal decomposition below in the results.

\section{Results}
\label{sec:Results}
\subsection{Spinodal Decomposition}

Spinodal decomposition is the process of relaxation of thermodynamically unstable states. Early studies on spinodal decomposition date back to the 1960s, with
the pioneering works of Hillert \cite{HILLERT}, Cahn and Hilliard
\cite{Cahn and HILLARD}, and Lifshiftz and Slyozov \cite{LIFSHITZ}.
Such studies were mainly focused on the mechanism of phase formation and macroscopic phase separation in
solid binary alloys and fluid binary mixtures.

At present it is well known that spinodal decomposition in binary mixtures has
three different regimes \cite{BRAY}. Initially some modulations present in the homogeneous phase grow
exponentially with time, following the early linear evolution dynamics.  With time the nonlinear coupling between these growing modes
slows their growth. In this second stage, the pattern has well-defined interfaces delimiting domains of the different stable
phases. At longer times the different domains of each phase
coalesce, due to interactions between the interfaces, in the so-called Ostwald ripening. The asymptotic state is decomposed into
two macroscopic domains, one for each phase.

In block copolymers, most of the studies of pattern formation and ordering investigate the
later stage of spinodal decomposition (the coarsening stage), focusing on the kinetic exponents of evolution, and comparison with
other related systems, like the self-assembled structures observed in Rayleigh-Benard convection.

Here we consider the early stages of spinodal decomposition in sphere-forming block copolymer systems under confinement in monolayers. We have observed two leading factors that control the degree of order during annealing below the spinodal. For $\tau>\tau_s$ the system is phase-separated but disordered and the kinetics of ordering are completely inhibited, while for $\tau \gtrsim \tau_s$ the strong length scale selectivity observed in Fig.  \ref{fig:unstablemodes} leads to well-defined hexagonal patterns with a lower density of defects and faster ordering kinetics. The differences in mode selectivity as a function of $\tau$ can be visualized in Fig. \ref{fig:cds3Da}, where  it can be appreciated that deep quenches lead to patterns with poorly defined symmetry.

In order to explore the dynamics and to facilitate comparison with the experimental results, here we have computed the circularly averaged structure factor for both experiments and numerical results.
The circularly averaged structure factor is defined as:
\begin{equation}
S_k = \langle \tilde{\psi}(\textbf{k})\, \tilde{\psi
^{*}}(\textbf{k}) \rangle
\end{equation}
where $\tilde{\psi} (\textbf{k})$ is the Fourier transform of the
order parameter.

For the experimental data we have calculated $S_k$ through the phase fields obtained by imaging the block copolymer with AFM. Although the density maps obtained by AFM are correlated to the local elasticity of the system, rather than to density fluctuations, they describe adequately the main pattern features: symmetry, defects, and degrees of local and long-range order.

\begin{figure}[b]
\begin{center}
\includegraphics[width=7.8cm]{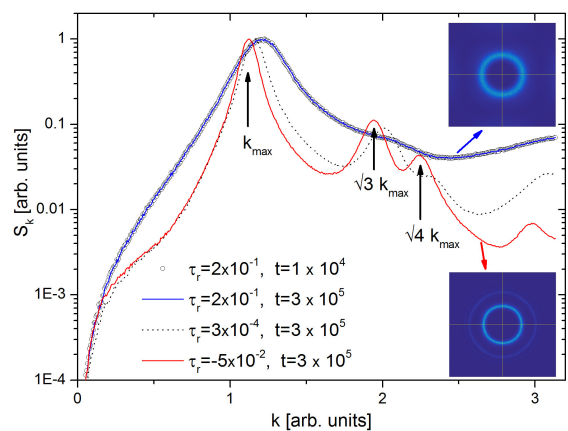}
\caption{Circularly averaged scattering function $S_k$ from simulations,  at different depths of quench and annealing times. The insets show the 2D patterns of $S_k$. The higher-order peaks, located at $\sqrt{3}k_{max}$, $\sqrt{4}k_{max}$ and  $\sqrt{7}k_{max}$, are the signature of hexagonal order.}
\label{fig:Skteo}
\end{center}
\end{figure}

Figure \ref{fig:Skteo} shows $S_k$ for systems quenched  and annealed at different conditions, as obtained from the simulations. Here, if the high temperature phase is deeply quenched into the spinodal region ($\tau_r=\tau /\tau_s-1=0.2$), the length scale selectivity imposed by the radius of gyration of the molecule leads to a nanophase-separated system with poorly defined symmetry. The insets in Fig. \ref{fig:Skteo} show the 2D scattering function for this system under two different conditions. The broad halo of intensity is consistent with a poorly defined symmetry and liquid-like order. In this case, there is not only one mode which can grow, but  rather a continuous range of modes delimited by an annulus determined by $k_1 \leq k \leq k_2$ (see Fig. \ref{fig:unstablemodes}). In this case, as the temperature is low and a wide spectrum of modes is stable, the kinetics of coarsening are frozen. Observe in Fig. \ref{fig:Skteo} that even after  $3$x$10^5$ time steps $S_k$ remains unaffected when $\tau_r=0.2$. Similar results were found by Yokojima
and Shiwa, and by Sagui and Desai, in a related system \cite{SHIWA1,SAG1,SAG2}. By
including thermal fluctuations it has been shown that the system can move towards equilibrium via defect annihilation and grain growth
\cite{VegaPRE,GomezPHYSD}.

When the same system is subjected to a shallow quench, $\tau_r=3$x$10^{-4}$, $S_k$ shows the typical features of hexagonal patterns, with a sharper main peak at $k_{max}$ and  well-defined higher-order peaks at $\sqrt{3}k_{max}$,  $\sqrt{4}k_{max}$ and $\sqrt{7}k_{max}$. The rings of nearly uniform intensity are also a signature of isotropy. However, in this case the isotropy emerges as a consequence of the polycrystalline structure and not a liquid-like order, as in the case of systems deeply quenched below $\tau_s$.

\begin{figure}[b]
\begin{center}
\includegraphics[width=7.7cm]{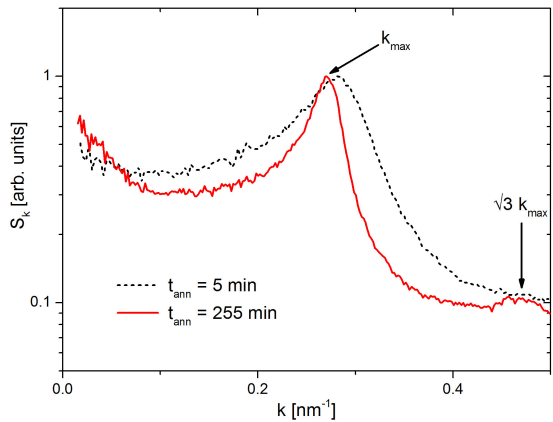}
  \caption{Circularly averaged scattering function $S_k$ for the experimental data at two different annealing times. Annealing temperature: T=333 K. Note that as annealing time increases, the main peak sharpens, shifts towards lower values of $k$, and develops a second-order peak at $\sqrt{3}k_{max}$, characteristic of a hexagonal pattern.}
\label{fig:Skexp}
\end{center}
\end{figure}

The dominant features of the scattering function during spinodal decomposition are in good qualitative agreement with the experimental data shown in Fig. \ref{fig:Skexp}. In the experimental system the early patterns are characterized by a strong length scale selectivity, but short-range order (see also Fig.  \ref{fig:fig1}). As the annealing time increases, there is a shift of the main peak in $S_k$ towards lower values of $k$. In addition, the main peak sharpens, and a weak higher-order peak located at $\sqrt{3}k_{max}$ can be clearly identified. These features are in good agreement with the results of Figs. \ref{fig:fig1} and \ref{fig:correxp}, which show an increasing degree of order with the annealing time.

Although there is qualitative agreement between experiments and numerical data,
it must be emphasized that the model employed here cannot capture the detailed dynamics of the system as a function of temperature. In block copolymers, the chain mobility depends on an effective monomeric friction coefficient that is strongly dependent on temperature, glass transition and/or crystallization temperatures of the individual blocks as well as the degree of segregation between blocks, the symmetry of the self-assembled structure, and the degree of entanglement. Although some of these properties can be qualitatively captured through an effective mobility coefficient ($M$ in Eqn. \ref{ecu:CHC}), at present it is not clear how these factors affect the dynamics.

\subsection{Nucleation and growth}
When the high-temperature phase is quenched to temperatures slightly above the spinodal (quenches with $\tau
\lesssim \tau_s$), all the modes decrease
exponentially in time and the order parameter goes to zero across the
whole system. That is, the initially randomly distributed order parameter $\psi(\textbf{r})$, characterized by $\langle\psi(\textbf{r})\rangle=0$ and $\langle\psi(\textbf{r})\psi(\textbf{r'})\rangle \ne 0$, dies off. However, in the same range of temperatures, an initial crystalline hexagonal
pattern is completely stable, indicating  a region of metastability $\tau_s<\tau<\tau_{ODT}$, where $\tau_{ODT}$ denotes the limit of metastability.

In addition, within this metastable region, a polycrystalline structure can improve its degree of order via the annihilation of defects, with a similar mechanism to the one observed for quenches only slightly below the spinodal. Figure \ref{fig:Skteo} also shows $S_k$ for a system quenched into this metastable region ($\tau_s>\tau>\tau_{ODT}$, $\tau_r=-5$x$10^{-2}<0$). Note that at the same time of annealing ($3$x$10^5$ timesteps), as compared with the system quenched below the spinodal ($\tau_r=3$x$10^{-4}>0$),  this system shows an improved order, as the main peak sharpens and higher-order peaks become better defined.

\begin{figure}[t]
\begin{center}
    \includegraphics[width=7.7cm]{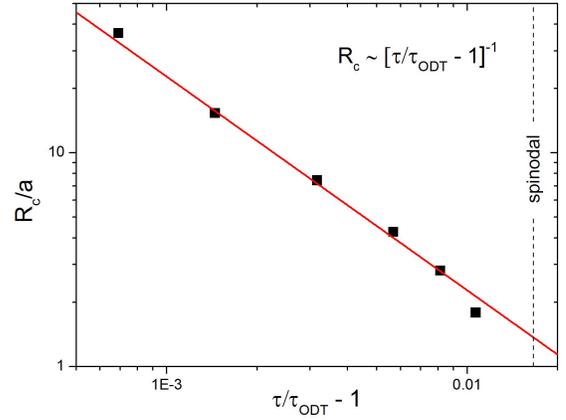}
    \caption{Critical nucleus size $R_c$ vs. the reduced temperature $\tau/\tau_{ODT}-1$. Here $a$ is the lattice constant. Note that $R_c \sim a$ for $\tau \sim \tau_s$.}
    \label{fig:radiocritico}
\end{center}
\end{figure}

The most distinctive feature of the metastable region is that only the nuclei or grains above a critical size can grow to develop the equilibrium phase, while smaller nuclei collapse due to the surface free energy.   According to the classical picture of nucleation and growth in 2D, the variation
in the free energy $\Delta F$ due to the formation of a nucleus of radius $R$
can be expressed as \cite{ChaikinBook,DebenedettiBook}:

\begin{equation}
\Delta F = 2 \sigma \pi R - 4 \pi |\Delta F_0| R^2
\end{equation}
where $\sigma$ represents the line tension and $\Delta F_0  \propto (\tau_{odt}-\tau) $ is the difference in the local free energies of the high-temperature (disordered) and the low-temperature (ordered) phases, which drives the transition. Due to the competition between these surface and volume contributions, only those
nuclei whose size exceeds a critical value $R_c$ can propagate to form the equilibrium phase.

\begin{figure}[t]
\begin{center}
\includegraphics[width=7.7cm]{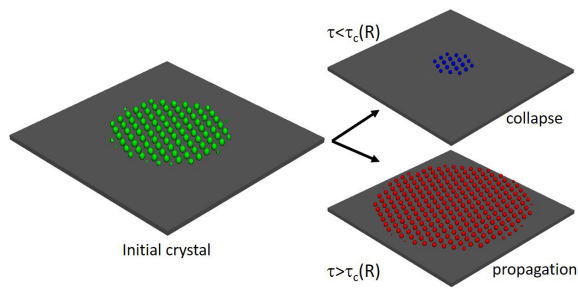}
    \caption{Time evolution of a crystalline domain as a function of temperature for a system quenched to $\tau_{odt}<\tau<\tau_{s}$. Here the left panel corresponds to a hexagonal grain of initial size $R_0/a=4.6$, where $a$ is the lattice constant.
     The right panels correspond to a grain that collapses (top, $\tau>\tau_c(R)$) or propagates (bottom, $\tau<\tau_c(R)$). }
     \label{fig:Rc}
\end{center}
\end{figure}

In order to obtain the dependence of the critical size for nucleation $R_c$ on the depth of quench, we explore the stability of crystalline nuclei of different sizes.
Figure \ref{fig:radiocritico} shows the critical size for nucleation $R_c$ as a function of $\tau$. Around the order-disorder
temperature $\tau_{ODT}$ we found that the critical size for grain growth $R_c$
diverges as $R_c \sim 1/(\tau/\tau_{ODT}-1)$, in agreement with
classical theories of nucleation and growth \cite{DebenedettiBook}. Note also that as the spinodal temperature is approached, the
critical size for nucleation becomes on the order of the lattice constant $a$. This result indicates that the process of spinodal decomposition could be inhibited, as the nucleation and growth process precedes spinodal decomposition \cite{DebenedettiBook}.

Figure  \ref{fig:Rc} shows typical snapshots of the evolution of nuclei as a function of temperature for quenches within the metastable region ($\tau_{odt}<\tau<\tau_s$). For an initial crystalline nucleus of a given size $R$, and with $R_c \propto (\tau_{odt}-\tau)^{-1}$, if the temperature is lowered to a value where $R$ is larger than $R_c$, grain growth occurs. Otherwise, the crystal collapses since line tension overcomes the bulk contribution.

\subsection{Melting}
A perfect crystalline structure may get trapped in a superheated state, such that the system may remain ordered when annealing above $\tau_{ODT}$.

\begin{figure}[t]
\begin{center}
    \includegraphics[width=7.7cm]{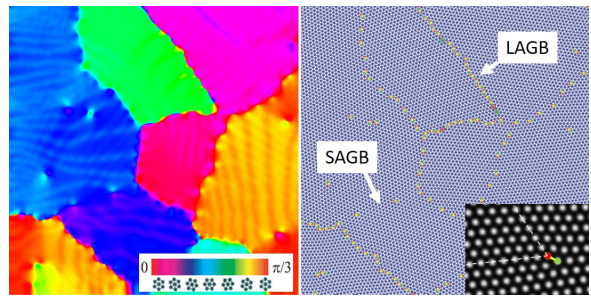}
    \caption{Orientational map $\theta(\textbf{r})$ (left panel) and defect structure (right panel) of the hexagonal lattice determined  by a Delaunay triangulation. Spheres with seven neighbors are indicated with a green dot,
those with five neighbors in red.  Dislocations are
formed by a pair of 5-7 disclinations separated by one lattice constant,
and are indicated by a connecting yellow line segment (see also the inset with a dislocation). A comparison between the defect structure and local orientation indicates that grain boundaries are decorated with linear arrays of dislocations.  The bottom of the figure for $\theta(\textbf{r})$ shows the color scale used to indicate the local orientation. Here the patterns correspond to $\tau_r=1.64$x$10^{-2}$ and  $9$x$10^4$ time steps.}
\label{fig:melt1}
     \end{center}
\end{figure}

\begin{figure}[th]
    \includegraphics[width=7.7cm]{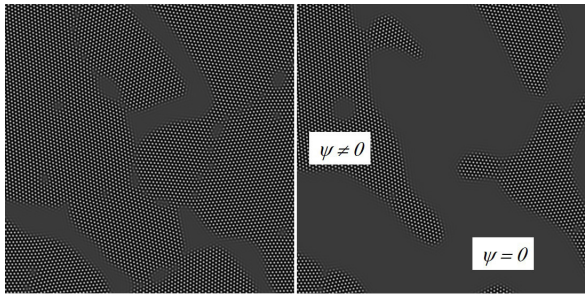}
    \caption{Upon increasing the temperature above $T_{ODT}$, the crystals of Fig. 10 melt first at LAGB (left panel). As time proceeds, the liquid phase ($\psi=0$) propagates through the system. $\tau=0.938\tau_s $. Left panel: $t=1.5$x$10^3$ time steps. Right panel: $t=5$x$10^4$ time steps.}
     \label{fig:melt2}
\end{figure}

However, we observed that for polycrystalline structures, topological defects trigger the phase transition, inhibiting superheating and disordering the system when $\tau \gtrsim \tau_{ODT}$.
Figure 10 shows a polycrystalline structure obtained during annealing at $\tau_r=1.64$x$10^{-2}$.  This figure shows the local orientation of the different hexagonal grains and also the defect structure.
In 2D crystals with hexagonal symmetry the elementary defects, named disclinations, are the domains which have a number of neighbors different from six (a five-coordinated domain is a positive disclination and a seven-coordinated domain is a negative disclination). In general, these defects are too energetic to be found in isolation: they couple in pairs to form dipoles, also known as dislocations (see Fig. \ref{fig:melt1}). Note in Fig. \ref{fig:melt1} that dislocations are piled up in linear arrangements, decorating the grain boundaries.
Figure \ref{fig:melt2} shows two snapshots of the system during annealing at temperatures above $T_{ODT}$.  Note in this figure the strong correlation between the liquid-like phase ($\psi=0$) and the position of the initial grain boundaries. In addition, it can also be observed that melting does not start uniformly at all grain boundaries, but depends on the orientational mismatch between neighboring crystals.

\begin{figure}[th]
    \includegraphics[width=7.7cm]{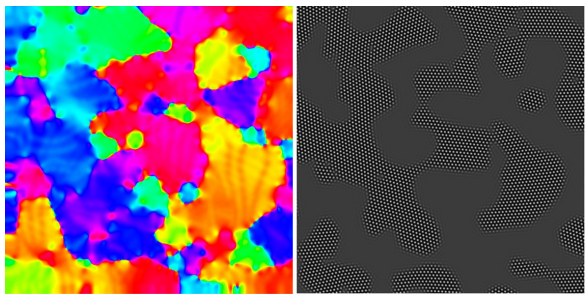}
    \caption{Melting of the crystal phase for a system containing a relatively large content of topological defects, shown initially at left. Although the temperature and annealing time are the same as in the left panel of Fig. \ref{fig:melt2}, due to the larger initial content of defects, there is a larger fraction of liquid phase here. $\tau=0.938\tau_s $, $t=1.5$x$10^3$ time steps.}
     \label{fig:melt3}
\end{figure}

The average distance between the dislocations located along a grain boundary depends on the orientation mismatch between neighboring grains, $\Delta \theta$, as  $d_{ds}\propto a/\Delta \theta^{-1}$, where $a$ is the lattice constant. Note that the largest possible mismatch in orientation between crystals in a hexagonal pattern is $30^{\circ}$.
Then, as compared with small-angle grain boundaries (SAGB), large-angle grain boundaries (LAGB) (those for which the orientational mismatch is in the range $10^{\circ}-30^{\circ}$), are more energetic and contain a larger number of dislocations per unit length. Consequently, the phase transition is initially triggered  at LAGB, where the crystal is more disordered.
We also observe that the kinetics of the order-disorder transition are affected by the initial defect structure. Note in Fig. \ref{fig:melt3} that a polycrystal with small crystal size, and thus a higher content of dislocations, leads to a larger fraction of the liquid phase at the same annealing time.

\section{Conclusion}
\label{sec:summary}

In this work we studied the different mechanisms leading to the
self-assembly of sphere-forming block copolymer thin films.
Using a Cahn-Hilliard free energy model valid for diblock copolymers, we followed the
evolution of the system with numerical simulations for different temperatures, and presented a
unified view of the phase transition process. The results are consistent with a first-order transition. Below a spinodal
temperature $\tau_{s}$, the initially homogeneous disordered state relaxes
towards equilibrium by spinodal decomposition, by the spontaneous growth of characteristic modes. Above this
temperature the system can only self-assemble into an ordered phase by nucleation and growth ($\tau_s < \tau < \tau_{ODT}$). Above the order-disorder temperature $\tau_{ODT}$ a well-ordered system can in principle remain superheated without disordering. However, dislocations and grain boundaries trigger the melting of the structure, such that defective states cannot be superheated above $\tau_{ODT}$.

For systems quenched into the metastable region, we found that the critical size
for nucleation and growth diverges as the disorder-order temperature is approached, following the law: $R_c \sim
1/(\tau/\tau_{ODT}-1)$, in agreement with classical theories of nucleation and growth.

\section*{Acknoledgment}

We gratefully acknowledge financial support from the National Science Foundation MRSEC Program through the
Princeton Center for Complex Materials (DMR-1420541), Universidad Nacional del Sur and the National Research Council of Argentina (CONICET).  The PS-PEP block copolymer was hydrogenated by Dr. John Sebastian.

\end{document}